\documentclass[12pt]{article} 
\usepackage{amssymb,amsmath,amsfonts}

\newcommand {\ca}       {{\mathfrak a}}

\newcommand {\alb}      {{\alpha}} 
\newcommand {\albb}     {{\bar\alpha}} 
\newcommand {\bal}      {{\bar\alpha}} 
\newcommand {\Fock}     {{\mathcal F}} 
\newcommand {\bFock}    {\bar{\mathcal F}} 
\newcommand {\fone}     {\varphi_{B1}} 
\newcommand {\ftwo}     {\varphi_{B2}} 
\newcommand {\bsl}[1]   {\langle {\mathfrak x}^{(u,v)}(#1)|} 
\newcommand {\bsr}[1]   {|{\mathfrak x}^{(u,v)}(#1)\rangle}

\newcommand {\depundeux}{x,\{b_k\},\{a_k\}} 
\newcommand {\depun}    {x_2,\{a_k\}} 
\newcommand {\depdeux}  {x_1,\{b_k\}} 
\newcommand {\noudep}   {\tilde x, \{c_k\}} 
 
\title{Boundary States for a Free Boson\\ 
Defined on Finite Geometries} 
\author{ 
Marc-Andr\'e Lewis\footnote{Email: {\ttfamily lewism{\char'100}cirano.umontreal.ca}}\\
Centre de recherches math\'ematiques, Montr\'eal, Canada, and\\
Laboratoire de Physique Th\'eorique et des Hautes Energies, \\
Universit\'es Pierre et Marie Curie (Paris VI) et Denis Diderot (Paris VII),\\
Paris, France\\
\and 
Yvan Saint-Aubin\footnote{Email: {\ttfamily saint{\char'100}crm.umontreal.ca}}\\ 
Centre de recherches math\'ematiques and\\ 
D\'ept.\ de math\'ematiques et de statistique,\\ 
Universit\'e de Montr\'eal, C.P.\ 6128, succ.\ centre-ville,\\ Montr\'eal, 
Qu\'ebec, Canada\ \ H3C 3J7 
}
\date{\today}

\begin{document} 
\maketitle 
 
\begin{abstract} 
Langlands recently constructed a map $\varphi\rightarrow 
|\mathfrak x(\varphi)\rangle$ that factorizes the 
partition function of a free boson on a cylinder with 
boundary condition given by two arbitrary functions 
$\varphi_{B_1}$ and $\varphi_{B_2}$ in the form 
$\langle\mathfrak x(\varphi_{B1})|q^{L_0+\bar L_0}|\mathfrak x(\varphi_{B2}) 
\rangle$. We rewrite $|\mathfrak x(\varphi)\rangle$ in a compact 
form, getting rid of technical assumptions necessary in his 
construction. This simpler form allows us to show explicitly 
that the map $\varphi\rightarrow |\mathfrak x(\varphi)\rangle$ 
commutes with conformal transformations preserving the 
boundary and the reality condition on the field $\varphi$. 
\end{abstract}

\noindent\hskip\leftmargin{\scshape Short Title:}{} Boundary States for a Free Boson

\noindent\hskip\leftmargin{\scshape PACS numbers:}{} 11.25.Hf, 11.30.-j, 05.30.Jp .

\noindent\hskip\leftmargin{\scshape Keyworks:}{} Boundary states, free boson,
free boson on a cylinder, conformal field theory, conformal invariance.

\section{Introduction} 
 
Renormalization transformations are often defined on 
spaces of fields or parameters of statistical physics
models. It is assumed that the existence of a non-trivial fixed point
of these transformations requires that the space it acts on be 
infinite-dimensional. Or at least that physical relevance of such
fixed points stems from the infinite number of degrees of
freedom. Langlands \cite{Dua} introduced a family of {\it finite}
models inspired by percolation, each endowed with a renormalization transformation 
with a non-trivial fixed point. Numerical analysis shows that, 
already for the coarsest models in the family, the critical
exponents bear some similarities with those accepted in the
literature for percolation.

The calculation presented here is a step towards extending
Langlands' construction for percolation to other models
with interaction. In \cite{RPL} Langlands calculated the partition
function of the free boson on a cylinder with fixed boundary
conditions. This partition function may be interpreted as the
probability of measuring given restrictions of the boson
at the cylinder extremities. Langlands argued that the space
of such probability distributions might be a natural set upon
which to construct a renormalization transformation. He showed
moreover how to construct vectors $|\mathfrak x_B\rangle$ in the
Fock space that represent the restriction at a given extremity
$B$ in such a way that the partition function is simply the expectation
value of the evolution operator between the two boundary states:
\begin{equation}
Z(\varphi|_{B_1},\varphi|_{B_2})=\langle \mathfrak x_{B_1}|q^{L_0\oplus \bar L_0}
|\mathfrak x_{B_2}\rangle.
\end{equation}
(The notation will be clarified in the following.) This is nothing
but the Feynman-Kac formula.
The present paper goes back to this formula to make Langland's
expression for the state $|\mathfrak x_{B}\rangle$
more explicit.  We should underline that it
is a rather unusual use of the Feynman-Kac formula.  While most of its
applications involve the computation of a partition function from the
evolution operator and boundary conditions, we use it as a tool to
define boundary states in a chosen algebraic structure.  The exercice is
more than aesthetical; using this new expression we can show that the 
boundary states transform properly under conformal transformations that
stabilize the boundary.

\section{Notations}

The description of free bosons is based on the Heisenberg algebra  
and its representations. The generators are 
the creation ($\ca_{-k}, \,k>0$), annihilation ($\ca_{k},\,k>0$) and central  
($\ca_0$) operators, obeying the commutation rule 
\begin{equation} 
[\ca_n,\ca_m] = n \delta_{n+m,0}. \label{u1c} 
\end{equation} 
The Fock space $\Fock_\alpha$ is a highest weight representation.  
The action of the generators on the highest weight vector $|\alpha\rangle$ 
is given by: 
\begin{align} 
\ca_{k} |\alpha\rangle &= 0, \qquad \forall k>0, \\ 
\ca_0 |\alpha\rangle &= \alpha |\alpha\rangle, 
\end{align} 
and physical states are generated by polynomials in the $\ca_{-k},\,k>0$.   
A basis for $\mathcal F_\alpha$ is given by the vectors 
\begin{equation}\label{base} 
|\alpha;n_1,n_2,\cdots\rangle = \ca_{-1}^{n_1}\ca_{-2}^{n_2} \cdots |\alpha\rangle, 
\end{equation} 
where the non-negative integers $n_i$'s are all zero but finitely many.  
The inner product on $\Fock_\alpha$ is  
defined by 
\begin{align} 
\langle \alpha'; n_{1}',n_{2}',\cdots| \alpha; n_1,n_2, \cdots\rangle &= \langle  
\alpha'| \cdots \ca_{3}^{n_3'}\ca_{2}^{n_2'}\ca_1^{n_1'} \ca_{-1}^{n_1}\ca_{-2}^{n_2} 
\ca_{-3}^{n_3}\cdots|\alpha\rangle\nonumber\\  
&= \delta_{\alpha,\alpha'}\left(\prod_{k=1}^\infty k^{n_k}\, n_k!\,  
\delta_{n_k,n_k'}\right). 
\end{align} 
The elements (\ref{base}) can thus be easily normalized to form an orthonormal 
basis. 
 
The Hilbert space of the free boson is the direct sum of tensor products 
of the form $\Fock_\alpha\otimes\Fock_\bal$, $\Fock_\alpha$ and $\Fock_\bal$ 
characterizing modes in the holomorphic and anti-holomorphic sectors,  
respectively. States in these tensor products are generated by the action of  
polynomials in the $\ca_{-k}$ and $\bar\ca_{-k}$ 
on the highest weight  
vector $|\alpha;\bar\alpha\rangle = |\alpha\rangle \otimes |\bar\alpha\rangle$.   
The generators $\ca_{k}$ are understood to act as $\ca_{k}\otimes 1$ and the  
$\bar\ca_k$ as $1\otimes \ca_k$. 
 
Fock spaces are given the structure of a Virasoro  
module by defining the conformal generators  
\begin{eqnarray} 
L_n &=& \frac{1}{2}\sum_{m\in\mathbb Z} :\ca_{n-m}\ca_m: \qquad n\ne 0 \nonumber \\ 
L_0 &=& \sum_{n>0} \ca_{-n}\ca_n + \frac{1}{2} \ca_0^2.    \label{Sug} 
\end{eqnarray} 
The expression for $L_0$ implies that ${\mathcal F}_\alpha$ is a  
highest weight module with highest weight $\alpha^2/2$.   
As will be described in the next section, the boson field is  
to be compactified on a circle of radius $R$. The pairs ($\alpha$, $\bar\alpha$) 
are then restricted to take the values 
\begin{align} 
\alpha = \alpha_{u,v} = \left(\frac{u}{2R} + vR\right) &\qquad\rightarrow 
\qquad h_{u,v} = \frac{1}{2}\left(\frac{u}{2R}+vR\right)^2\\ 
\bar\alpha = \bar\alpha_{u,v} = \alpha_{u,-v} &\qquad\rightarrow\qquad  
\bar h_{u,v} = \frac{1}{2}\left(\frac{u}{2R}-vR\right)^2, 
\end{align} 
with $u$ and $v$ integers and where $h_{u,v}$ and $\bar h_{u,v}$ are the values 
of $L_0=L_0\otimes 1$ and $\bar L_0=1\otimes L_0$ acting on 
$\mathcal F_{\alpha_{u,v}}\otimes \mathcal F_{\bar\alpha_{u,v}}$. 
We will denote $\mathcal F_{\alpha_{u,v}}$ 
by $\mathcal F_{(u,v)}$ and $\mathcal F_{\bar\alpha_{u,v}}$ 
by $\bar{\mathcal F}_{(u,v)}$. 
 
In his calculation, Langlands chose the Virasoro algebra Vir as the fundamental 
structure. He was able to construct explicitly the map $\mathfrak x$  
for irreducible Verma modules over Vir. 
However Verma modules over Vir are reducible for some 
rational highest weights. (Rational  
compactification radii $R$ do lead to such highest weights.) 
It was his suggestion that we look for an alternative definition 
that would encompass reducible cases. Using the Heisenberg algebra 
as the basic structure avoids this difficulty and leads as well 
to an elegant form for $\mathfrak x$.

\section{Explicit calculation of the partition function}

We identify the cylinder with the quotient of the infinite strip  
$0<\mbox{Re}\,w<\ln \,q$, $0<q<1$, by the translations $w\rightarrow  
w+2\pi ik,\,k\in\mathbb Z$.  It can be mapped on the annulus $\mathcal A$  
of center 0, outer radius 1 and inner radius $q$ by the conformal map $z=e^w$.   
The angle $\theta$ 
of the annular geometry parametrizes both extremities of the cylinder. 
 
The partition function is defined as 
\begin{equation} 
\int {\mathcal D}\varphi\, e^{-\int_{\mathcal A} {\mathcal L}(\varphi) d^2 z} 
\end{equation} 
where $\int_{\mathcal A}$ denotes the integration over the annulus,  
and the Lagrangian density is given by 
\begin{equation} 
{\mathcal L}(\varphi) = \partial_z \varphi \partial_{\bar z} \varphi. 
\end{equation} 
The usual mode expansion of $\varphi(z,\bar z)$ is 
$$\varphi(z,\bar z)=\varphi_0+a \ln z+ b \ln\bar z+\sum_{n\neq 0}\left( 
\varphi_n z^n + 
\bar\varphi_n \bar z^n\right).$$ 
The restriction $\varphi_{B_1}$ of this field to the 
inner circle where $z=q e^{i\theta}$ and $\bar z=q e^{-i\theta}$ is of 
the form: 
$$\varphi_{B1}(\theta) = \varphi_0+(a+b) \ln q+i\theta(a-b) 
+\sum_{k\ne 0} b_k e^{ik\theta}, \qquad b_{-k}=\bar b_k$$ 
and the restriction $\varphi_{B_2}$ 
to the outer circle ($z=e^{i\theta}, \bar z=e^{-i\theta}$): 
$$\varphi_{B2}(\theta) = \varphi_0+i\theta(a-b) 
+\sum_{k\ne 0} a_k e^{ik\theta}, \qquad a_{-k}=\bar a_k.$$ 
(The relationship between $a_k, b_k$ and $\varphi_n$ will 
be given below.) 
Since it is the field $e^{i\varphi/R}$ that really matters, $\varphi$ 
need not be periodic but should only satisfy the milder requirement 
$\varphi(e^{2\pi i}z, e^{-2\pi i}\bar z)=\varphi(z,\bar z)+2\pi vR$, 
$v\in\mathbb Z$. This statement is equivalent to the compactification of the 
field $\varphi$ on a circle of radius $R$ and implies that  
$$a-b=-ivR,\qquad v\in\mathbb Z.$$ 
The Lagrangian density does not depend on $\varphi_0$ and this constant 
may be set to zero. Therefore only the difference of the 
constant terms in $\varphi_{B_1}$ and $\varphi_{B_2}$ remains. 
We choose to parametrize this difference 
by a real number $x\in [0,2\pi R)$ and 
an integer $m\in\mathbb Z$: 
$$-(a+b)\ln q=x+2\pi mR.$$ 
The reason for this parametrization is again the compactification 
of $\varphi$: even though the various pairs  
$(\varphi_{B_1}+2\pi mR,\varphi_{B_2})$, $m\in\mathbb Z$, will give different  
contributions to the functional integral, they all represent 
the same restriction of $e^{i\varphi/R}$ at the boundary. 
 
We are interested in computing the partition function $Z(\varphi_{B_1}, 
\varphi_{B_2})=Z(x,\{b_k\},\{a_k\})$ defined as 
\begin{equation} 
Z(x,\{b_k\},\{a_k\}) = \int_{B} {\mathcal D}\varphi\, e^{-\int_{\mathcal A}  
{\mathcal L}(\varphi) d^2 z}, 
\end{equation} 
where $\int_B$ denotes the integration on the  
space of functions $\varphi$ such that the restrictions 
of $e^{i\varphi/R}$ at the inner and outer boundaries coincide 
with $e^{i\varphi_{B_1}/R}$ and $e^{i\varphi_{B_2}/R}$. 
(The dependence on the compactification radius $R$ is always implicit.) 
The decomposition of the field in a classical part verifying the
boundary conditions and fluctuations vanishing at the
extremities leads to 
\begin{equation} 
Z(x,\{b_k\},\{a_k\}) = \Delta^{-1/2} Z_{\text{class}}(x,\{b_k\},\{a_k\}).
\end{equation} 
The factor $\Delta$ is the $\zeta$-regularization of the determinant for the  
annulus and is known to be (see for example \cite{RPL,MSD}): 
\begin{equation} 
\Delta^{-1/2} = (-i\tau)^{-1/2} \eta^{-1}(\tau),\qquad\text{with\ } q=e^{i\pi\tau}, 
\end{equation} 
where $\eta(\tau)= e^{i\pi\tau/12} \prod_{m=1}^{\infty} (1-e^{2im\pi\tau})$ is  
the Dedekind $\eta$ function. The factor $Z_{\text{class}}$ 
is the integration (sum) over all classical solutions 
compatible with the boundary conditions in the above sense. 
To obtain $Z_{\text{class}}$ we solve the classical 
equations ($\partial_z\partial_{\bar z}\varphi=0$) with boundary 
conditions given by $(\varphi_{B_1},\varphi_{B_2})$. The condition at 
the outer circle ($z=e^{i\theta}, \bar z=e^{-i\theta}$) is $\varphi_n+ 
\bar\varphi_{-n}=a_n$ and that at the inner one ($z=qe^{i\theta}, 
\bar z=qe^{-i\theta}$) is 
$q^n\varphi_n+q^{-n}\bar\varphi_{-n}=b_n$. The solution can be written 
as the sum  
\begin{equation}\label{modefourier} 
\varphi=a\ln z+b\ln\bar z+\tilde\varphi_1+\tilde\varphi_2 
\end{equation} 
where the two function $\tilde\varphi_1$ and $\tilde\varphi_2$ 
are harmonic inside the annulus and take respectively the 
values $\fone$ and $0$ on the inner boundary and the values $0$ and $\ftwo$  
on the outer one.  These functions are 
$$\tilde\varphi_{1}(z,\bar z) = \sum_{k\ne 0} \frac{b_k}{q^k-\frac{1}{q^k}}  
(z^k - \bar z^{-k}),$$ 
$$\tilde\varphi_{2}(z,\bar z) =  \sum_{k\ne 0} \frac{a_k}{\frac{1}{q^k}-q^k}  
\left(\left({\frac{z}{q}}\right)^k - \left({\frac{\bar z}{q}}\right)^{-k}\right).$$ 
Hence the classical solution $\varphi$ is  
completely determined by the data $(x,\{b_k\},\allowbreak 
\{a_k\})$ up to the two integers $m,v\in\mathbb Z$ that determine $a$ and $b$. 
The factor $Z_{\text{class}}$ is consequently the sum: 
$$\sum_{m,v\in\mathbb Z} e^{\mathcal L(\varphi_{(m,v)})}$$ 
where $\varphi_{(m,v)}$ is the solution (\ref{modefourier}) 
with $-(a+b)\ln q=x+2\pi mR$ and 
$a-b=-ivR$. 
 
Using this expression and the Poisson summation formula on the index $m$,  
Langlands \cite{RPL} computed the desired partition function as the product 
\begin{equation}\label{Zexp} 
Z(x,\{b_k\},\{a_k\}) = \Delta^{-1/2} Z_1(x) Z_2(\{b_k\},\{a_k\}) 
\end{equation} 
where 
\begin{equation}\label{robert} 
Z_1(x)=\sum_{u,v\in\mathbb Z} e^{iux/R}q^{\frac{u^2}{4R^2}+v^2R^2}= 
\sum_{u,v\in\mathbb Z} e^{ix(\alpha_{u,v}+\bar\alpha_{u,v})}\, 
q^{h_{u,v}+\bar h_{u,v}} 
\end{equation} 
and 
\begin{equation} 
Z_2 (\{b_k\},\{a_k\}) = 
\prod_{k=1}^\infty\exp\left(-2k\left( 
  \frac{1+q^{2k}}{1-q^{2k}} (a_ka_{-k} + b_kb_{-k}) 
- \frac{2q^k}{1-q^{2k}}     (a_kb_{-k} + b_ka_{-k}) 
\right)\right). 
\end{equation}

\section{Explicit form of the boundary states}

In this section we rewrite (\ref{Zexp}) as a sum over $u,v\in\mathbb Z$ 
of terms of the form  
\begin{equation} 
Z^{(u,v)}(\depundeux) =  \bsl{\fone} q^{L_0+\bar L_0} \bsr{\ftwo}, 
\label{explicite} 
\end{equation} 
with $\bsr{\varphi} \in \Fock_{(u,v)} \otimes \bFock_{(u,v)}$ 
and where  
\begin{equation*} 
Z^{(u,v)}(\depundeux) =\Delta^{-\frac12}e^{iux/R} 
q^{\frac{u^2}{4R^2}+v^2R^2} Z_2 (\{b_k\},\{a_k\}). 
\end{equation*} 
The goal is therefore 
to find a map $\mathfrak x^{(u,v)}$ such that (\ref{explicite}) holds. 
To do so we will first set $Z_2 (\{b_k\},\{a_k\})$ in the form 
\begin{equation} 
Z_2 (\{b_k\},\{a_k\}) = \prod_{k=1}^\infty \sum_{m,n} B^k_{m,n}  
q^{k(m+n)} A^k_{m,n}\label{but} 
\end{equation} 
where $A^k_{m,n}=A^k_{m,n}(a_k,a_{-k})$ and  
$B^k_{m,n}=B^k_{m,n}(b_k,b_{-k})$ are functions of only two variables. 
With $Q=q^k$, $a_\pm = 2i\sqrt{k} a_{\pm k}$ and $b_\pm = -2i\sqrt{k}  
b_{\pm k}$ for $k\ge 1$, the terms $\Delta^{-1/2} Z_2$ of (\ref{Zexp})  
are a product over $k$ of 
\begin{equation} 
e^{a_+a_-/2}e^{b_+b_-/2} \frac{e^{(a_+a_-Q^2 +a_+b_-Q+b_+a_-Q+b_+b_-Q^2) 
/(1-Q^2)}}{1-Q^2}, \label{aplus} 
\end{equation} 
up to a constant depending only on $\tau$. 
The factors in front are clearly factorizable and can be absorbed in the  
definition of $A^k_{m,n}$ and $B^k_{n,m}$.  The remaining mixed term can  
be developped as a power series in $Q$: 
\begin{equation} 
\sum_{i,j,k,l=0}^\infty \frac{(a_+a_-+b_+b_-)^i}{i!} 
\frac{(a_+b_-)^j(b_+a_-)^k}{j!k!} 
\frac{(i+j+k+l)!}{(i+j+k)!l!}Q^{2i+j+k+2l}. \label{dev} 
\end{equation} 
To achieve the form (\ref{but}), the coefficient of the term 
$q^{k(m+n)}$ (i.e.\ of $Q^{2i+j+k+2l}$ with $2i+j+k+2l=m+n$) 
in the above expression has to be the product of two functions, 
one of $(a_k,a_{-k})$, the other of $(b_k,b_{-k})$. We 
concentrate on the terms with $j\ge k$ and denote by $S_{m,n}$ 
the factor of $(a_+b_-)^{m-n}Q^{m+n}$ with $j-k=m-n$. The terms 
with $j<k$ are treated similarly. With the use of  
$x=a_+a_-$ and $y=b_+b_-$, $S_{m,n}$ can be written as 
$$S_{m,n} (x,y) = \sum_{i+k+l=n} \frac{(x+y)^i(xy)^k}{i!(m-n+k)!k!} 
\frac{(m+k)!}{(m-n+i+2k)!l!}.$$ 
This function is clearly symmetric in $x$ and $y$.  Define 
$$R_{m,n}(x) \equiv S_{m,n}(x,0)$$ 
and 
$$T_{m,n} \equiv R_{m,n}(0) = S_{m,n}(0,0).$$ 
Casting (\ref{dev}) in the form (\ref{but}) will only be possible if 
\begin{equation} 
S_{m,n}(x,y)T_{m,n} = R_{m,n}(x)R_{m,n}(y). \label{condnec} 
\end{equation} 
That this condition is verified is highly non-trivial.  It was found  
in \cite{RPL} that it is, although in disguised form, the Saalsch\"utz  
identity \cite{GKP}.  We thus have the factorization if we define 
\begin{align} 
A^k_{m,n}(a_k,a_{-k}) = \frac{R_{m,n}(x)}{\sqrt{T_{m,n}}} 
a_+^{m-n}e^{a_+a_-/2},\qquad \text{if\ }m\ge n 
\end{align} 
where  
$$R_{m,n}=\sum_{i=0}^n \frac{m!x^i}{i!(m-n)!(m-n+i)!(n-i)!},\qquad 
m\ge n,$$ 
and 
$$T_{m,n}=\frac{m!}{n!((m-n)!)^2},\qquad m\ge n.$$ 
This expression for $R_{m,n}$ shows that it is related to the 
$n$-th Laguerre polynomial of the $(m-n)$-th kind by 
\begin{equation} 
R_{m,n}(x) =\frac {L_n^{(m-n)}(-x)}{(m-n)!},\qquad m\ge n. 
\end{equation} 
A similar calculation leads to  
\begin{equation} 
R_{m,n}(x)=\frac{L^{(n-m)}_m(-x)}{(n-m)!},\qquad n\ge m. 
\end{equation} 
Going back to the initial notation,  
we finally get the desired form with: 
\begin{equation} 
A^k_{m,n}(a_k,a_{-k}) =\begin{cases} 
(2i\sqrt{k} a_k)^{m-n} \sqrt{\frac{n!}{m!}}e^{-2k|a_k|^2}  
L_n^{(m-n)} (4k|a_k|^2) & m \ge n  \\ 
(2i\sqrt{k} a_{-k})^{n-m} \sqrt{\frac{m!}{n!}} e^{-2k|a_k|^2}  
L_m^{(n-m)} (4k|a_k|^2) & n \ge m,  
\end{cases} 
\end{equation} 
and $B_{m,n}^k (b_{k},b_{-k})=\overline{A^k_{m,n}(b_{-k},b_{k})}$. (An  
orientation on the boundary must be chosen to define the 
map ${\mathfrak x}$. For example moving in the positive direction 
of the parameter $\theta$ should put the cylinder at one's left. This 
explains the interchange $b_k\leftrightarrow b_{-k}$ in the functions 
$B$.) 
 
It is now straightforward to define the map from the boundary conditions  
to the Hilbert space.  We have just shown that the contribution  
of the $(u,v)$ sector to the partition function can be written as 
$$Z^{(u,v)}(\depundeux) = q^{h_{u,v} + \bar h_{u,v}} 
e^{ix(\alpha_{u,v}+\bar\alpha_{u,v})}  
\prod_{k\in\mathbb N} \sum_{m,n=0}^\infty  
B^k_{m,n} q^{k(m+n)} A^k_{m,n}.$$ 
(Note that both sectors $(u,v)$ and $(u,-v)$ contribute the same quantity to 
$Z$. There seems therefore to be a freedom to attach $(u,v)$ to either  
$\mathcal F_{(u,v)}\otimes \bar{\mathcal F}_{(u,v)}$ or  
$\mathcal F_{(u,-v)}\otimes\bar{\mathcal F}_{(u,-v)}$.  
This choice is resolved in the next section.)   
Using the fact that 
$$\langle u,v|\frac{(\ca_k^m\otimes\ca_k^n)\,(\ca_{-k'}^{m'} \otimes  
\ca_{-k'}^{n'})}{\sqrt{k^{m+n}{k'}^{m'+n'} m!n!m'!n'!}}|u,v\rangle =  
\delta_{k,k'} \delta_{m,m'} \delta_{n,n'},$$ 
we get 
\begin{align} 
Z^{(u,v)}&(\depundeux) =\\ 
&=q^{h_{u,v} + \bar h_{u,v}} e^{ix(\alpha_{u,v}+\bar\alpha_{u,v})}  
\prod_{k,k'} \sum_{m,m',n,n'} B^k_{m,n} q^{k(m+n)} A^{k'}_{m',n'} 
\delta_{k,k'}\delta_{m,m'}\delta_{n,n'}\\ 
&=\langle {\mathfrak x}^{(u,v)}(\depdeux) | q^{L_0\oplus\bar L_0}|  
{\mathfrak x}^{(u,v)}(\depun)\rangle 
\end{align} 
where 
\begin{equation}|{\mathfrak x}^{(u,v)}(\depun)\rangle =  
e^{ix_2(\alpha_{u,v}+\bar\alpha_{u,v})}  
\prod_{k=1}^\infty \sum_{m,n=0}^\infty A^k_{m,n}(a_k,a_{-k}) 
\frac{\ca_{-k}^m \otimes \ca_{-k}^n}{\sqrt{k^{m+n}m!n!}}  
|u,v\rangle.\label{ef}\end{equation} 
We have reintroduced, somewhat arbitrarily, the constant term 
$x_2$ in $\varphi_{B2}$. Again, only the difference $x=x_2-x_1$ 
between the constant term $x_2$ in $\varphi_{B2}$ and $x_1$ in 
$\varphi_{B1}$ has a physical meaning. 
We now have an explicit form for the map ${\mathfrak x}$.   
 
The vector $|\mathfrak x^{(u,v)}(\depun)\rangle$ can be cast into 
a simpler form. With the help of the following recursion identities 
$$(n+1)L^{(m-(n+1))}_{n+1}(x)-[x\partial_x -x+(m-n)]L^{(m-n)}_n(x)=0, 
\qquad m-1\ge n\ge 0$$ 
and 
$$L^{((m+1)-n)}_n(x)+[\partial_x-1]L^{(m-n)}_n(x)=0,\qquad m\ge n\ge 0,$$ 
we can prove by induction on both indices that 
\begin{equation} 
A^k_{m,n}=\frac{(\partial_- + \frac{1}{2} a_+)^m(\partial_+ +\frac{1}{2} a_-)^n} 
{\sqrt{m!n!}} e^{a_+a_-/2}, 
\end{equation} 
where, we recall, $a_\pm = 2i\sqrt{k} a_{\pm k}$ and $\partial_{\pm}  
= \frac{\partial}{\partial{a_\pm}}$. 
Defining 
$\alpha_{k} = \frac{i}{2} (-\partial_{-k} + 2k a_{k}),\, 
\bar\alpha_{k} = \frac{i}{2} (-\partial_{k} + 2k a_{-k})$ 
and $\Omega_{k} = A^k_{0,0}=e^{a_+a_-/2}=e^{-2k|a_k|^2}$, 
we get  
\begin{equation} 
A^k_{m,n} = \frac{\alpha_{k}^m 
\bar\alpha_{k}^n}{\sqrt{k^{m+n}m!n!}} \Omega_{k}. 
\end{equation} 
The correspondence ${\mathfrak a}_{-k}\leftrightarrow i\alpha_k$ 
and $\bar{\mathfrak a}_{-k}\leftrightarrow i\bar\alpha_k$ induces 
an isomorphism with a subalgebra of the Heisenberg algebra since 
the $\alpha_k$'s and $\bar\alpha_k$'s satisfy the  
commutation rules: 
$$[\alpha_{n},\alpha_{m}] = -n\delta_{n+m,0} \qquad  
[\bar\alpha_{n},\bar\alpha_{m}] = -n\delta_{n+m,0}$$ 
$$[ \alpha_n,\bar \alpha_{m}] = 0.$$ 
If $|\alpha_{uv}\rangle\otimes|\bar\alpha_{u,v}\rangle$ is identified 
with $\Omega=\prod_k\Omega_k$ and $\alpha_0$ (resp.\ $\bar\alpha_0$) 
is defined as acting by multiplication by $\alpha_{u,v}$ (resp.\ 
$\bar\alpha_{u,v}$), this correspondence can then be extended to an 
isomorphism of Heisenberg modules. 
Since the $\ca_{-k}$'s and $\alpha_k$'s, $k>0$, all commute 
with one another, we are able to write down  
an exponential form for the boundary state: 
\begin{eqnarray} 
\bsr{\depun} &=& e^{ix_2(\alpha_{u,v}+\bar\alpha_{u,v})}\prod_{k}\left\{ 
 \sum_{m,n}  \frac{(\alb_k\ca_{-k})^m (\albb_k\bar\ca_{-k})^n}{ 
m!n! k^m k^n} \Omega_k \right\}|u,v \rangle\nonumber\\ 
   &=& e^{ix_2(\alpha_{u,v}+\bar\alpha_{u,v})}\prod_{k} \left 
     \{ e^{\alb_k\ca_{-k}/k} e^{\albb_k\bar\ca_{-k}/k}\right\}  
     \Omega  |u,v \rangle\nonumber \\  
   &=&  e^{ix_2(\alpha_{u,v}+\bar\alpha_{u,v})}\prod_{k}  \left 
     \{e^{(\alb_k\ca_{-k}+\albb_k\bar\ca_{-k})/k}\right\}  
     \Omega  |u,v \rangle\nonumber\\ 
   &=& e^{ix_2(\alpha_{u,v}+\bar\alpha_{u,v})} 
 e^{\sum_{k\in\mathbb N} (\alb_k\ca_{-k}+\albb_k\bar\ca_{-k})/k}  
     \Omega  |u,v \rangle.\label{fonc} 
\end{eqnarray} 
Up to the factor $(-i\tau)^{-1/2}e^{-i\pi\tau/12}$ the partition function 
takes the following form: 
\begin{equation} 
Z(\depundeux) = \langle {\mathfrak x}(\depdeux)| q^{L_0+\bar L_0}  
| {\mathfrak x}(\depun) \rangle, 
\end{equation} 
in which we have defined 
\begin{equation}\label{lareponse} 
|{\mathfrak x}(\depun)\rangle = e^{ix_2(\ca_0+\bar\ca_0)}  
e^{\sum_{k=1}^\infty (\alb_k\ca_{-k}+\albb_k\bar\ca_{-k})/k}\Omega |\Lambda\rangle 
\end{equation} 
\begin{equation} 
|\Lambda\rangle = \bigoplus_{u,v} |u,v\rangle, 
\end{equation} 
where the operators $\ca_0=\ca_0\otimes 1$ and $\bar\ca_0=1\otimes 
\ca_0$ act as the identity times $\alpha_{u,v}$ and $\bar\alpha_{u,v}$ 
on $|u,v\rangle=|\alpha_{u,v}\rangle \otimes |\bar\alpha_{u,v}\rangle$. 
The boundary states $|{\mathfrak x}(\depun)\rangle$ 
belong to the direct sum of Fock spaces $\bigoplus_{u,v}  
\Fock_{(u,v)}\otimes\bFock_{(u,v)}$ or, more precisely, to the sum  
$\bigoplus_{u,v} (\Fock_{(u,v)}\otimes\bFock_{(u,v)})^c$ of some completions  
that contains 
formal series like (\ref{lareponse}). 
In the next section, it will turn out to be useful to include the 
$x$-dependence in $\Omega$, which will then be noted 
$\Omega_{u,v}$, to highlight its sector: 
\begin{equation} 
\Omega_{u,v} = e^{ix(\alpha_{u,v}+\bar\alpha_{u,v})}\Omega. 
\end{equation} 

The boundary state $|{\mathfrak x}(\depun)\rangle$ has an analogue
in string theory. The propagation of a closed string can be
wrtitten as the expectation value of an evolution operator
between in and out states. In light-cone coordinates, these states
are labeled by Fourier coefficients of $(D-2)$ periodic coordinate
functions, $D$ being the spacetime dimension, and their expression
has been given in \cite{Sie,Nepo}. Restricting the expressions
(12, 13) in \cite{Nepo} to a single coordinate function, the 
dependency on the Fourier coefficients $a_k, k\neq 0$, can be
shown (after a somewhat lengthy calculation) to be identical
to ours. Both expressions, theirs and ours, have an overall
phase that depends on the zero mode $a_0$. Their phase does
not depend on winding numbers but ours does. The next section
will show that, for the problem at hand, the phase in 
$|{\mathfrak x}(\depun)\rangle$ is crucial to assure the proper
behavior of $|{\mathfrak x}\rangle$ under conformal transformations
in each Fock sector $\Fock_{(u,v)}\otimes\bFock_{(u,v)}$.

\section{Conformal Transformations and Boundary States}

Having found an explicit and concise form for the boundary states,  
we can now study their properties under conformal transformations.   
Let $g$ be an infinitesimal 
conformal transformation that leaves the boundary unchanged 
and $G$ the corresponding element in the Virasoro algebra.  
The 
purpose of this section is to show that the action of $g$ on 
the boundary condition $\varphi$ and that of $G$ on $|{\mathfrak x} 
(\varphi)\rangle$ commute: 
\begin{equation} 
|{\mathfrak x}(g\varphi)\rangle = G|{\mathfrak x}(\varphi)\rangle. \label{prop} 
\end{equation} 
We first discuss the actions $g$ and $G$ and 
the correspondence between them.
 
One can easily convince oneself that the only infinitesimal conformal 
transformations that preserve the center and radius  
of a circle in the complex plane are linear combinations of 
\begin{equation} 
(l_p-\bar l_{-p}), \qquad p\in\mathbb Z, 
\end{equation} 
where the conformal generators $l_p$ and $\bar l_p$ are  
defined as  $l_p = -z^{p+1} \partial_z$ and $\bar l_p = 
-\bar z^{p+1} \partial_{\bar z}$. Note that the subalgebra  
$\oplus_{p\in\mathbb Z}\mathbb C(L_p-\bar L_{-p})\subset \text{Vir}\otimes 
\overline{\text{Vir}}$ is centerless and the mapping defined by  
$(l_p-\bar l_{-p})\rightarrow (L_p-\bar L_{-p})$ of the boundary preserving 
conformal transformation into $\text{Vir}\otimes\overline{\text{Vir}}$ 
is an isomorphism. However the tranformations $(l_p-\bar l_{-p}), p\neq 0$, 
do not preserve the reality condition imposed on the boundary functions.   
The generators $(l_p+\bar l_p)$ and $i(l_p-\bar l_p)$ do.  
Both reality and geometry preserving conditions are therefore satisfied by  
the infinitesimal transformations  
\begin{eqnarray} 
g_0^{(1)} &=& 1+i\epsilon\left\{l_0-\bar l_0\right\} \\ 
g_p^{(1)} &=& 1+i\epsilon\left\{(l_p+ l_{-p}) -  
(\bar l_p + \bar l_{-p})\right\}, \qquad p > 0 
\end{eqnarray} 
and 
\begin{equation} 
g_p^{(2)} = 1+\epsilon\left\{(l_p- l_{-p}) +  
(\bar l_p - \bar l_{-p})\right\}, \qquad p>0.\end{equation} 
We shall show that (\ref{prop}) holds if the $g_p^{(i)}$'s are defined as  
above and the corresponding $G_p^{(i)}$'s are taken to be 
\begin{eqnarray} 
G_0^{(1)} &=& 1 + i \epsilon\{L_0 - \bar L_0\}, \\ 
G_p^{(1)} &=& 1+i\epsilon\left\{ (L_p+ L_{-p}) - 
(\bar L_p + \bar L_{-p})\right\}
\end{eqnarray} 
and 
\begin{equation} 
G_p^{(2)} = 1+\epsilon\left\{ (L_p- L_{-p}) + (\bar L_p - \bar L_{-p})\right\}. 
\end{equation} 
Since, for $p\ne 0$, we have 
$$ 
[g_p^{(1)}-1,g_0^{(1)}-1]= -\epsilon p (g_p^{(2)}-1), 
$$ 
the property for the second family of transformations follows directly  
if it is proven to be true for the first one. The action of $G$ in the rhs of  
(\ref{prop}) is simply left-multiplication. On the lhs, the action is defined 
as usual by $(g\varphi)(z,\bar z)=\varphi \circ g^{-1} (z, \bar z)$. 
We first study the case  
$p>0$. For $p=0$, the  
particularity of the Sugawara construction will modify  
the analysis. We will end this section by examining this case.   
 
Let us first compute $|{\mathfrak x}(g_p\varphi)\rangle$ with $g_p=g_p^{(1)}$.  
Note that, due to the use of the Poisson summation formula to  
obtain (\ref{robert}), the constant $m$ in $-(a+b)\ln q=x+2\pi mR$ is not 
anymore well-defined in the sector $(u,v)$. However the difference  
$(a-b)$ still is. Only $a-b$ will appear 
in the variation $g_p\varphi$. As observed in the previous 
paragraph, the two contributions $Z^{(u,v)}$ and $Z^{(u,-v)}$ 
are equal. It turns out that equation (\ref{prop}) holds when the 
functions $\varphi$ with a given $v$ are mapped into the sectors 
$\mathcal F_{(u,v)}\otimes \bar{\mathcal F}_{(u,v)}$, 
$u\in\mathbb Z$. (For the other choice  
$\mathcal F_{(u,-v)}\otimes \bar{\mathcal F}_{(u,-v)}$, 
the actions $g$ and $G$ fail to commute.) 
The function on the boundary must have the 
form 
$$\varphi(\theta)=x+vR\theta+\sum_{k> 0}(a_ke^{ik\theta}+a_{-k} 
e^{-ik\theta})$$ 
or, equivalently 
$$\varphi(z,\bar z)=x+(a\ln z+b\ln \bar z)+\sum_{k>0} 
(a_kz^k+\bar a_k\bar z^k)$$ 
with $z=e^{i\theta}$ and $\bar z=e^{-i\theta}$ and the reality 
condition $a_{-k}=\bar a_{k}$. A direct calculation gives 
$$g_p\varphi=\tilde x+vR\theta + \sum_{k>0} c_k e^{ik\theta}+\bar c_{k} 
e^{-ik\theta}$$ 
where 
\begin{align} 
c_k      &= a_k + i\epsilon\left((k+p)a_{k+p} + (k-p)a_{k-p}\right)  
   + \epsilon vR\delta_{k,p} \label{ckak}\\ 
\bar c_k &= \bar a_k - i\epsilon\left((k+p)\bar a_{k+p} +  
(k-p)\bar a_{k-p}\right) 
   + \epsilon vR\delta_{k,p} \label{ckakb}\\ 
{\tilde x}&= x + i\epsilon p(a_{p} - \bar a_{p}).\label{xtx} 
\end{align} 
One can see that the reality condition imposed on $\varphi$  
is indeed preserved.  Since 
\begin{align} 
|{\mathfrak x}^{(u,v)}(\noudep)\rangle &=\left.\left(e^{\sum_{k>0}  
({\alpha}_k\ca_{-k}+{\bar\alpha}_k\bar\ca_{-k})/k} {\Omega}_{u,v}\right)\right|_{ 
g_p\varphi} |u,v\rangle\\ 
&= e^{\sum_{k>0} (\tilde{\alpha}_k\ca_{-k}+ 
\tilde{\bar\alpha}_k\bar\ca_{-k})/k} \tilde{\Omega}_{u,v}|u,v\rangle 
\end{align} 
where we have defined  
\begin{align} 
\tilde{\alb}_k &=  \frac{i}{2}(-\frac{\partial}{\partial \bar c_k} + 2k c_k),\\ 
\tilde{\albb}_k &= \frac{i}{2}(-\frac{\partial}{\partial c_k}      + 2k \bar c_k),\\ 
{\tilde\Omega}_{u,v} &=  e^{i{\tilde x}(\alpha_{u,v}+\bar\alpha_{u,v})}  
e^{-2\sum_{k>0} k c_k \bar c_{k}}, 
\end{align} 
a first step is to express $\tilde\alb_k$, $\tilde\albb_k$, and  
$\tilde\Omega_{u,v}$ in terms of $x$ and the $a_k$'s. This can 
be easily achieved. 
The expressions for $c_k$ and ${\bar c}_k$ given 
above can be inverted in order to obtain closed 
form expressions for $a_k$ and ${\bar a}_k$. 
It is then a simple exercise to show that    
\begin{align*} 
\tilde{\alb}_k &= \frac{i}{2}(-\frac{\partial}{\partial \bar c_k} + 2kc_{k})  
=\begin{cases}\alb_k + i\epsilon k(\alb_{k+p}+\alb_{k-p}),& k\neq p,\\ 
\alb_p + i\epsilon p(\alpha_{2p} + \alpha_{uv}),& k=p, 
\end{cases}\\ 
\tilde{\albb}_k &= \frac{i}{2}(-\frac{\partial}{\partial c_k} + 2k\bar c_{k})  
=\begin{cases}\albb_k - i\epsilon k(\albb_{k+p}+\albb_{k-p}),& k\neq p,\\ 
	      \albb_k - i\epsilon p(\albb_{2p} + \albb_{uv}),& k=p. 
\end{cases} 
\end{align*}

Using these expressions, the functional  
$\tilde\Omega_{u,v}=\left.\Omega_{u,v}\right|_{g_p\varphi}$  
can be expressed in terms of the original variables. A careful treatment 
of the infinite sums leads to 
\begin{equation} 
{\tilde\Omega}_{u,v}=\left(1+i\epsilon\Big(\frac{1}{2}\sum_{0<k<p} 
(\alpha_{p-k}\alpha_k - \bar\alpha_{p-k}\bar\alpha_k) +  
\alpha_p\alpha_{u,v} - \bar\alpha_p\bar\alpha_{u,v}\Big)\right)\Omega_{u,v}. 
\end{equation} 
 
Finally we can rewrite $\bsr{g_p\varphi}$ to 
first order in $\epsilon$ as 
\begin{align} \label{eq57}
\bsr{g_p\varphi} &= e^{\sum_k (\tilde{\alpha}_k \ca_{-k} +  
  {\tilde{\bar\alpha}}_k\bar\ca_{-k})/k} \tilde{\Omega}_{u,v}|u,v\rangle \nonumber\\ 
         &=e^{\sum_k(\alpha_k\ca_{-k} + \albb_k\bar\ca_{-k})/k} \nonumber\\ 
         & \times\left(1+i\epsilon \sum_{k>0} \Big( 
         (\alpha_{k+p}+\alpha_{k-p})\ca_{-k} - 
         (\bar\alpha_{k+p}+\bar\alpha_{k-p})\bar\ca_{-k}\Big)\right.\nonumber \\ 
& \left.\,\,\,+ i\epsilon\Big(\alpha_p{\mathfrak a}_0 -  
\bar\alpha_p\bar{\mathfrak a}_0+\frac{1}{2}\sum_{0<k<p} 
(\alpha_{p-k}\alpha_k - \bar\alpha_{p-k}\bar\alpha_k) \Big)\right)\Omega_{u,v}|u,v\rangle. 
\end{align} 
 
We now turn our attention to the rhs of (\ref{prop}), namely $G_p| 
{\mathfrak x}(\varphi)\rangle$.  It is convenient to introduce the operator 
$u^k_{m,n}$ defined as 
\begin{equation} 
u^k_{m,n} =\frac{ {\mathfrak a}_{-k}^m\otimes  
{\mathfrak a}_{-k}^n }{\sqrt{k^{m+n} m!n!}}. 
\end{equation} 
It is such that 
\begin{equation} 
\langle u,v | u_{m,n}^{k\ \dagger} |{\mathfrak x}(\varphi)\rangle =  
A^k_{m,n}(\varphi) \frac{\Omega_{u,v}}{\Omega_k},\label{xgphi} 
\end{equation} 
where, we recall, $\Omega_k=e^{-2k|a_k|^2}$. Moreover 
\begin{equation} 
\left(u^{k'}_{m',n'}|u,v\rangle\right)^\dagger  
\left(u^k_{m,n}|u,v\rangle\right) = \delta_{k,k'}\delta_{m,m'}\delta_{n,n'} 
\end{equation} 
and 
\begin{equation}\label{complet} 
\prod_{k=1}^\infty \sum_{m,n=0}^\infty u^{k}_{m,n}|u,v\rangle\langle u,v|u^{k\ \dagger}_{m,n} 
\end{equation} 
acts as the identify on $\mathcal F_{(u,v)}\otimes\bar{\mathcal F}_{(u,v)}$. 
 
From the Sugawara construction 
\begin{equation} 
L_{p} = \frac{1}{2} \sum_{k\in\mathbb Z} :\mathfrak a_{p-k} \mathfrak a_k:, \qquad p\ne 0, 
\end{equation} 
we see that 
\begin{equation} 
[L_{p},{\mathfrak a}_{k}^m] = -mk \ca_{k}^{m-1}{\mathfrak a}_{k+p} 
\end{equation} 
with similar equations for the anti-holomorphic sector.  
Also, since $\langle u,v|\ca_{k} =0$ if $k<0$, 
\begin{eqnarray} 
\langle u,v | L_{p} &=& \langle u,v |\Big(\frac{1}{2} 
\sum_{0<k<p} {\mathfrak a}_k {\mathfrak a}_{p-k} + {\mathfrak a}_p  
{\mathfrak a}_0\Big). 
\end{eqnarray} 
For the same reason, the operator $L_{-p}$ annihilates  
$\langle u,v|$ when acting from the right.  We thus have 
\begin{eqnarray} 
\langle u,v |u^{k\ \dagger}_{m,n} L_{p} &=& \langle u,v |\left([u^{k\ \dagger}_{m,n},L_p]  
+   L_p u^{k\ \dagger}_{m,n}\right) \nonumber\\ 
      &=& \langle u,v | \left( mk\frac{({\mathfrak a}_k^{m-1}\ca_{k+p})  
      \otimes {\mathfrak a}_k^n}{\sqrt{k^{m+n}m!n!}} + L_p u^{k\ \dagger}_{m,n}\right) 
\end{eqnarray} 
and 
\begin{eqnarray} 
\langle u,v |u^{k\ \dagger}_{m,n} L_{-p} &=& \langle u,v | \left([u^{k\ \dagger}_{m,n}, 
        L_{-p}]\right) \nonumber\\ 
      &=& \langle u,v | \left( mk\frac{({\mathfrak a}_k^{m-1}{\mathfrak a}_{k-p})  
        \otimes {\mathfrak a}_k^{n}}{\sqrt{k^{m+n}m!n!}}\right). 
\end{eqnarray} 
With these two relations and their equivalent in the anti-holomorphic sector,  
we can compute $\langle u,v |u^{k\ \dagger}_{m,n} G_p$: 
\begin{eqnarray} 
\langle u,v |u^{k\ \dagger}_{m,n} G_p &=& \langle u,v |\left(u^{k\ \dagger}_{m,n} + i\epsilon k  
\frac{\ca_k^{m-1}\otimes \ca_k^{n-1}}{\sqrt{k^{m+n}m!n!}}  
(m(\ca_{k+p}+\ca_{k-p})\bar\ca_k - n\ca_k  
(\bar\ca_{k+p}+\bar\ca_{k-p}))\right)\nonumber \\ 
     & & \,\, + i\epsilon\langle u,v |\left(
      {\mathfrak a}_p {\mathfrak a}_0  
     - \bar\ca_p\bar\ca_0+\frac{1}{2}\sum_{0<l<p}  
     {\mathfrak a}_l {\mathfrak a}_{p-l}- \bar{\mathfrak a}_l  
     \bar{\mathfrak a}_{p-l} \right)u^{k\ \dagger}_{m,n}. 
\end{eqnarray} 
We thus get 
\begin{align} 
\langle u,v |u^{k\ \dagger}_{m,n} G_p | {\mathfrak x}(\varphi)\rangle     
  &= \left(1+ i\epsilon\Big(
  \alpha_p \alpha_{u,v}-\bar\alpha_p\bar\alpha_{u,v}
  +\frac{1}{2}\sum_{0<l<p}  
  \alpha_l\alpha_{p-l}-\bar\alpha_l\bar\alpha_{p-l}   
  \Big)\right) 
  A^k_{m,n}\frac{\Omega_{u,v}}{\Omega_k} \nonumber\\ 
  & + i\epsilon k \frac{\alpha_k^{m-1}\bar\alpha_k^{n-1}} 
  {\sqrt{k^{m+n}m!n!}}\Big(m (\alb_{k+p}+\alpha_{k-p})\albb_{k} -  
  n \alb_k (\albb_{k+p}+\albb_{k-p})\Big)\Omega_{u,v}.  
\end{align} 
Once again $\bar\alpha_0$ acts on $\Omega_{u,v}$ by  
multiplication by $\bar\alpha_{u,v}$. 
Using the completeness relation (\ref{complet}), we can  
reconstruct $G_p\bsr{\varphi}$.  First, summing over $m$ and $n$ gives 
\begin{eqnarray} 
&&\sum_{m,n} \frac{\langle u,v |u^{k\ \dagger}_{m,n} G_p |  
 {\mathfrak x}^{(u,v)}(\varphi)\rangle}{\prod_{l\ne k}\Omega_l}  
 \, u^k_{m,n} |u,v\rangle \nonumber\\ 
&&\,\,= \frac{1}{\prod_{l\ne k} \Omega_l} \left(1+  
 i\epsilon\left(\frac{1}{2}\sum_{0<l<p} \alpha_l\alpha_{p-l} 
 -\bar\alpha_l\bar\alpha_{p-l}  + \alpha_p {\mathfrak a}_0 
 -\bar\alpha_p\bar\ca_0\right)\right)\left(\sum_{m,n}   
 A^k_{m,n}\prod_{l\ne k} \Omega_l \frac{\ca_{-k}^m\bar\ca_{-k}^n} 
 {\sqrt{k^{m+n} m!n!}} \right)|u,v\rangle \nonumber\\ 
&&\,\,\,\, + \frac{i\epsilon}{\prod_{l\ne k} \Omega_l} 
 \left(k(\alpha_{k+p}+\alpha_{k-p}) \sum_{m\ge 1,n\ge 0}  m  
 \alpha_k^{m-1}\bar\alpha_k^{n}\frac{\ca_{-k}^m\bar\ca_{-k}^n} 
 {k^{m+n}m!n!}\right)\Omega_{u,v}|u,v\rangle \nonumber\\ 
&&\,\,\,\, - \frac{i\epsilon}{\prod_{l\ne k} \Omega_l} 
 \left(k(\bar\alpha_{k+p}+\bar\alpha_{k-p}) \sum_{m\ge 0,n\ge 1}   
 n \alpha_k^{m}\bar\alpha_k^{n-1}\frac{\ca_{-k}^m\bar\ca_{-k}^n} 
 {k^{m+n}m!n!}\right)\Omega_{u,v}|u,v\rangle. \nonumber 
\end{eqnarray}  
The last two terms can be rewritten as 
\begin{eqnarray} 
\frac{i\epsilon}{\prod_{l\ne k} \Omega_l}\left((\alpha_{k+p}+\alpha_{k-p}) 
\ca_{-k} \sum_{m\ge 0,n\ge 0}  \alpha_k^{m}\bar\alpha_k^{n} 
\frac{\ca_{-k}^m\bar\ca_{-k}^n}{k^{m+n}m!n!}\right) 
\Omega_{u,v}|u,v\rangle \nonumber\\ 
- \frac{i\epsilon}{\prod_{l\ne k} \Omega_l} 
\left((\bar\alpha_{k+p}+\bar\alpha_{k-p})\bar\ca_{-k}  
\sum_{m\ge 0,n\ge 0}  \alpha_k^{m}\bar\alpha_k^{n} 
\frac{\ca_{-k}^m\bar\ca_{-k}^n}{k^{m+n}m!n!}\right)\Omega_{u,v}|u,v\rangle. 
\end{eqnarray} 
Putting all this together, we finally have 
\begin{align} 
G_p|{\mathfrak x}^{(u,v)}(\varphi)\rangle &= e^{\sum_k(\alpha_k\ca_{-k}  
+ \albb_k\bar\ca_{-k})/k} \nonumber\\ 
&\times\left[1+i\epsilon \sum_{k>0} \Big((\alpha_{k+p}+\alpha_{k-p}) 
 \ca_{-k} -(\bar\alpha_{k+p}+\bar\alpha_{k-p})\bar\ca_{-k}\Big)\right. \nonumber \\ 
&\left.\,+i\epsilon\Big(\alpha_p{\mathfrak a}_0-\bar\alpha_p 
 \bar{\mathfrak a}_0+\frac{1}{2}\sum_{0<k<p}(\alpha_{p-k}\alpha_k  
 - \bar\alpha_{p-k}\bar\alpha_k)\Big)\right]\Omega_{u,v}|u,v\rangle\nonumber\\ 
&= |{\mathfrak x}^{(u,v)}(g_p\varphi)\rangle. 
\end{align} 
We have thus established the desired property for $p>0$.  
It is the phase $e^{ix (\alpha_{uv}+\bar\alpha_{uv})}$
in $|{\mathfrak x}\rangle$ and the winding term $vR\theta$
in $\varphi$ that are responsible for the terms $\mathfrak
a_0$ and $\bar{\mathfrak a}_0$ in eq.\ (\ref{eq57}). Their
role is therefore crucial to prove the conformal
property (\ref{prop}).
 
The transformation $g_0^{(1)}=1+i\epsilon(l_0-\bar l_0)$  
is nothing but an infinitesimal rotation.  
The gaussian terms are invariant under these  
transformations, because Fourier coefficients only pick up a phase. Hence 
$\left.\Omega_{u,v}\right|_{g_0^{(1)}\varphi}=(1+i\epsilon vR 
(\alpha_{u,v}+\bar\alpha_{u,v}))\left.\Omega_{u,v}\right|_\varphi$. 
The computation of $\bsr{g_0^{(1)}\varphi}$ is straightforward and one gets 
\begin{eqnarray} 
|{\mathfrak x}^{(u,v)}(g_0^{(1)}\varphi)\rangle &=&e^{\sum_k(\alpha_k\ca_{-k}  
+ \albb_k\bar\ca_{-k})/k}\nonumber\\ 
&&\,\,\times \left(1+i\epsilon \Big(\sum_{k>0} \alpha_{k}\ca_{-k} -  
\bar\alpha_{k}\bar\ca_{-k}\Big) + i\epsilon vR(\alpha_{u,v}+\bar\alpha_{u,v}) 
\right) 
\Omega_{u,v}|u,v\rangle\nonumber\\ 
  &=&e^{\sum_k(\alpha_k\ca_{-k} + \albb_k\bar\ca_{-k})/k}\nonumber\\ 
&&\,\,\times \left(1+i\epsilon \Big(\sum_{k>0} \alpha_{k}\ca_{-k}  
-\bar\alpha_{k}\bar\ca_{-k}\Big)+ i\epsilon(h_{u,v} -  
\bar h_{u,v})\right)\Omega_{u,v}|u,v\rangle. 
\end{eqnarray} 
The action of $G_0$ is somehow different. 
In this case, the $\bar L_0$ term does not 
annihilate $\langle u,v|$, but rather acts  
on it by multiplying by $\bar h_{u,v}$.  We thus get 
\begin{eqnarray} 
G_0|{\mathfrak x}^{(u,v)}(\varphi)\rangle &=& e^{\sum_k(\alpha_k\ca_{-k} +  
   \albb_k\bar\ca_{-k})/k}\nonumber\\ 
 &&\,\,\times \left(1+i\epsilon \Big(\sum_{k>0} \alpha_{k}\ca_{-k}  
   -\bar\alpha_{k}\bar\ca_{-k}\Big)+ i\epsilon (h_{u,v}  
   - \bar h_{u,v})\right)\Omega_{u,v}|u,v\rangle\nonumber\\ 
 &=& |{\mathfrak x}^{(u,v)}(g_0^{(1)}\varphi)\rangle. 
\end{eqnarray} 
This completes the proof.

\section{Concluding remarks}

This simple yet quite instructive calculation gives an  
example of a conformal theory with non-conformally 
invariant boundary conditions. Can the map $\varphi\rightarrow 
|\mathfrak x(\varphi)\rangle$ for the free boson be used to 
investigate minimal models with general boundary conditions? 
It is well known that minimal models  
can be constructed from the $c=1$ CFT, using the Coulomb  
gas technique. This was succesfully done on the plane by Dotsenko and  
Fateev \cite{DF1,DF2} and on the torus by Felder \cite{Felder}.  
This might be one path to construct the map for these models. 

Langlands and the two authors have recently studied numerically
the statistical distribution of the Fourier coefficients of a field
defined for the Ising model. This distribution is more intricate
than the boson's as the Fourier coefficients of the field at one
boundary do not appear now to be mutually independent. The map
$\varphi\rightarrow 
|\mathfrak x(\varphi)\rangle$, if it exists for the Ising
model, might be a rich object.

\section*{Acknowledments} 
 
The authors would like to thank Robert Langlands for explaining his original 
construction and stressing its limitations, Philippe Zaugg for helpful 
discussions and Rafael Nepomechie for bringing to our attention
the papers in string theory relevant to the present work.
 
M.-A.\ L.\ gratefully acknowledges fellowships from the NSERC Canada  
Scholarships Program and the Celanese Foundation, and Y.\ S.-A.\ support 
from NSERC (Canada) and FCAR (Qu\'ebec).

\end{document}